\documentclass[prd,twocolumn,showpacs, preprintnumbers,showkeys]{revtex4}
\usepackage{graphicx}
\usepackage{amsfonts}
\begin{document}
\preprint{ANL-HEP-PR-08-06}

\title{Entanglement entropy in $d+1$ $SU(N)$ gauge theory}
\author{Alexander Velytsky}
\email{vel@theory.uchicago.edu}
\affiliation{Enrico Fermi Institute, University of Chicago, 5640 S. Ellis Ave., Chicago, IL 60637, USA,}
\affiliation{HEP Division and Physics Division, Argonne National Laboratory, 9700 Cass Ave., Argonne, IL 60439, USA}
\date{\today}

\begin{abstract}
We consider the entanglement entropy for a sub-system in $d+1$ dimensional $SU(N)$ lattice gauge theory. The $1+1$ gauge theory is treated exactly and shows trivial behavior. Gauge theories in higher dimensions are treated within Migdal-Kadanoff approximation. 
We consider the gauge theory in the confinement phase. We demonstrate the existence of a non-analytical change from the short distance to long distance form in the entanglement entropy in such systems ($d>2$) reminiscent of phase transition. The transition is manifested in nontrivial change in the RG flow of character expansion coefficients defining the partition function.
\end{abstract}
\maketitle

\section{Introduction}
Interest in the study of entanglement entropy has a relatively 
long history in quantum field theory. The early motivation was due to 
the connection with black hole physics. General properties of the 
entanglement entropy, such as its dependence only on the surface, 
were demonstrated for system of oscillators and massless non-interacting scalar field theory 
\cite{Srednicki:1993im}. Later the 
entanglement entropy was studied in gravity duals of confining large $N_c$ gauge theories 
\cite{Klebanov:2007ws} using the AdS/CFT 
approach of \cite{Ryu:2006bv}. 
In this work  the $d$ dimensional space was divided  into two complementary regions $A$ and $\bar{A}$ by two imaginary $d-1$ 
dimensional hyper-surfaces placed 
distance $l$ apart along one of the space directions
\begin{eqnarray}
A&=&\mathbb{R}^{d-1}\times \mathbb{I}_l,\nonumber\\
\bar A&=&\mathbb{R}^{d-1}\times (\mathbb{R}- \mathbb{I}_l),
\label{eq:region}
\end{eqnarray}
where $\mathbb{I}_l$ is a line segment of length $l$.
The authors studied the entanglement entropy as a function of $l$ and found that it exhibits a non-analytical change in behavior at $l=l^*_c$ 
reminiscent of a 
phase transition\footnote{Similar results were obtained for the static AdS bubble solution in \cite{Nishioka:2006gr}.}.

In the present work we aim to prove that this is a general scenario for $SU(N)$ gauge theories (at arbitrary $N$) at temperatures 
corresponding to the confinement phase.
We consider a $d+1$ dimensional gauge theory at finite temperature $T$. The zero temperature system is recovered as the limit $T
\rightarrow 0$.
We consider the same geometry of the entangled region as in \cite{Klebanov:2007ws}, see eq. (\ref{eq:region}). 

If we are given the density matrix for such a system, we can integrate out all degrees of freedom associated with region $\bar A$. The 
resulting density matrix can be used to construct the entanglement entropy
\begin{equation}
\rho_A={\rm Tr}_{\bar A} \rho, \quad S_A=-{\rm Tr}_A\rho_A \log \rho_A,
\end{equation}
which is the entropy as seen by an observer with no access to the degrees of freedom in $\bar A$.

We will use a method of gluing replicas of the system under consideration into a multi-sheet Riemann surface, which was used in an 
extensive treatment of 2d CFT in \cite{Calabrese:2005zw,Calabrese:2004eu}. We consider $n$ replicas of  a system after the trace over  
region $\bar A$ has been taken (for this the boundaries in time direction of this region were identified)\footnote{The boundary of region $
\bar A$ is treated according to a standard finite temperature field theory formalism.}. Each of these replicas is glued to another along the  
boundary of region $A$ normal to the time direction. The first replica's upper boundary (coordinate $t=1/T$) is glued to the lower 
boundary (coordinate $t=0$) of the second replica and so on. The upper boundary of the last $n$-th replica is glued to the lowest 
boundary of the first replica, thus closing the system. For illustration of such gluing in 1+1 and 2+1 dimensional theories see Figs. 
\ref{fig:zn2d} and \ref{fig:zn3d}. One can observe that in such a system 
\begin{equation}
{\rm Tr}\rho_A^n=\frac{Z_n(A)}{Z^n},
\end{equation}
where $Z_n$ is the partition function of the glued system and $Z$ is the standard partition function of the original system 
($Z=Z_1$).

This approach allows one to construct the entanglement entropy
\begin{equation}
S_A=-\lim_{n\rightarrow 1}\frac\partial{\partial n}{\rm Tr}\rho_A^n=
-\lim_{n\rightarrow 1}\frac\partial{\partial n}\frac{Z_n(A)}{Z^n}
\end{equation}

\section{$SU(N)$ gauge theory in $d+1$ dimensions}
The partition function for $SU(N)$ lattice gauge theory is
\begin{equation}
Z=\int\prod_ldU_l \prod_p e^{-S_p},
\label{eq:pf}
\end{equation}
where the action is
$S_p\equiv S(U_p)=-\beta/(2N){\rm Tr}U_p+h.c.$, $\beta=2N/g^2$ is the lattice inverse coupling, and the plaquette variable is the ordered 
product of gauge fields which live on the links constituting the plaquette 
$U_p=\prod_{l\in\partial p}U_l$. The gauge invariant action is a class function and therefore it can be expanded in group characters
\begin{equation}
e^{-S_p}=\sum_r F_r d_r\chi_r(U_p)\equiv F_0\left(1+\sum_{r\neq0}c_rd_r\chi_r(U_p)\right),
\label{eq:char}
\end{equation}
where the first sum runs over all irreducible representations, while the second sum excludes the trivial $r=0$ representation. For general 
$SU(N)$ group $r$ is a set of indices; $d_r$ is the dimension of the representation,
$c_r=F_r/F_0<1$ and $F_r$ are the coefficients of expansion
\begin{equation}
F_r=\int dU e^{-S(U)}\frac{1}{d_r}\chi^*_r(U).
\label{eq:fr}
\end{equation}

\subsection{$d=1$ gauge theory}
\begin{figure}[ht]
\centering
\includegraphics[width=0.55\columnwidth]{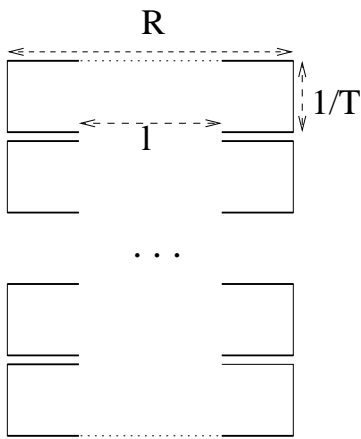}
\caption{$Z_n$ for $1+1$ dimensional gauge theory.}
\label{fig:zn2d}
\end{figure}

The 2-dimensional $SU(N)$ gauge theory is exactly solvable, see \cite{Gross:1980he} for an overview and large $N$ treatment of zero 
temperature $U(N)$ gauge theory.
It is possible to treat the zero temperature gauge theory in this study in analogous fashion, however we would like to consider a more 
general case of a gauge theory at finite temperature $T$ and therefore adopt a different approach. Finite temperature gauge theory in 2 
dimensions normally is formulated on a $\mathbb{R}\times \mathbb{S}_1$ surface periodic in time direction with period $1/T$. For 
practical reasons we consider a finite system in space direction. The corresponding discretized theory is formulated on a $N_r\times N_t$ 
lattice, with space-time cut-off $a$ and $a N_t=1/T$ and $aN_r=R$. 

At this point it is instructive to consider how integration on the surface is performed. Consider an elementary part of a surface bounded by 
a single loop\footnote{We define the elementary surface as the surface with all dynamical degrees of freedom belonging to the surface 
perimeter, with no degrees of freedom in the interior.}. Its contribution to the resulting partition function\footnote{Note that there is also a 
"bulk" contribution $F_0$, see (\ref{eq:char}).}  is
\begin{equation}
f(\{a\}; \partial A)\equiv 1+\sum_{i\neq0}d_ia_i\chi_i(\partial A),
\end{equation}
where $\partial A$ is a product of link variables along the surface perimeter and the function is defined when all $a_i$ coefficients are 
specified. The expression for a junction
of two surface elements $A$ and $B$ with a common boundary $A\cap B$ is
\begin{eqnarray}
f(\{c\}; \partial (A\cup B)) &=& \int d(A\cap B) f(\{a\}; \partial A) f(\{b\}; \partial B) \nonumber\\
&=&  1+\sum_{i\neq0}d_ic_i\chi_i(\partial (A\cup B)),\nonumber\\
c_i&=&a_ib_i.
\label{eq:join}
\end{eqnarray}
The integration over the common boundary $U=A\cap B$ 
is performed using the character
property: 
\begin{equation}
\int dU\chi_r(VU)\chi_s(U^\dagger W)=\frac1{d_r}\delta_{r,s}\chi_r(VW).
\end{equation}
In other words the junction of the surfaces in the space of character coefficients is represented by an ordinary product.

For any 2-dimensional surface we can expand  the partition function (\ref{eq:pf}) in characters according to (\ref{eq:char}) 
and then integrate all the internal plaquettes using (\ref{eq:join}). The resulting expression for the partition function is
\begin{equation}
Z=\int\prod_{l\in \partial A}dU_l  \sum_r F_r^A d_r\chi_r(U_{\partial A}),
\label{eq:pf2}
\end{equation}
where $A=N_r N_t$ is the area of the total surface in plaquette units (number of tiling plaquettes) and 
$\partial A$ is the contour enclosing the surface. 

The multi sheet n-replica partition function $Z_n$ for 2-d model is shown in Fig. \ref{fig:zn2d}. 
Remember that the links of time boundary of region $\bar A$ for each replica (bold lines) are identified and so are the links  from region 
$A$ that form the time boundary of $Z_n$ (dotted lines).
Being subjected to the same treatment as $Z$ the partition function of the glued system $Z_n$ will result in the same expression 
(\ref{eq:pf2}), but with corresponding surface area $A_n=n A=n N_r N_t$ and perimeter $\partial A_n$. 

To perform the perimeter integration first  we choose free boundary condition (b.c.) in the spatial direction. 
The invariance of the group integration (Hurwitz/Haar measure) allows one to manipulate the link variables in the perimeter integral, so 
that the final integration is performed over a single plaquette perimeter (c.f. Gross-Witten one plaquette integral) in both $\partial A$ and $
\partial A_n$ integrations. Specifically we absorb the space-like links that separate different replicas in $Z_n$ (time boundary of region $
\bar A$) so that the system becomes identical to a simple plaquette with free spatial b.c.

Due to periodicity in time direction the ordered contour product of gauge fields generally has the form
\begin{equation}
U_{\partial A}=U_{0,\hat 1}V_{1,\hat 0 }U_{0,\hat 1}^\dagger V_{2,\hat 0}^\dagger.
\end{equation}
Here $U_{n,\hat i}$ denotes the gauge field at coordinate $n$ in $\hat i=0,1$ direction, where $\hat0$ is chosen to be the time direction.

We use another property of character integration
\begin{equation}
\int dU_{0,\hat 1} \chi_r(U_{0,\hat 1}V_{1,\hat 0}U_{0,\hat 1}^\dagger V_{2,\hat 0}^\dagger)=\frac 1{d_r}\chi_r(V_{1,\hat 0})\chi_r(V_{2,\hat 
0}^\dagger)
\end{equation}

The integral over the remaining two gauge variables decouples and has support only for the trivial
representation  $\chi_0=1$. This leads to a simple result   
\begin{equation}
Z= F_0^A.
\end{equation}
The ratio of the partition functions is unity and the entanglement entropy is zero.

Next we consider a lattice periodic in the spatial direction 
which effectively mimics infinite spatial extent. 
The perimeter integral for $Z$ now is
\begin{equation}
\int dV\int dU \chi_r(UVU^\dagger V^\dagger)=\int dV \frac 1{d_r}\chi_r(V)\chi_r(V^\dagger)=\frac1{d_r},
\label{eq:zper}
\end{equation}
where in the last part we used the character orthonormality property.
The partition function becomes
\begin{equation}
Z= \sum_r F_r^A.
\end{equation}
It is easy to check that the $Z_n$ perimeter integral results in
\begin{eqnarray}
&&\int dU_1...dU_n\frac1d_r\frac{\chi_r(U_1)...\chi_r(U_n)}{d_r^{n-1}}\frac{\chi_r(U_1^\dagger)...\chi_r(U_n\dagger)}{d_r^{n-1}}\nonumber\\
&=&\frac1{d_r^{2n-1}}. 
\end{eqnarray}
Note that this expression for $n=1$ correctly reproduces the result of perimeter inegration for $Z$, c.f.  (\ref{eq:zper}).
The partition function ratio is
\begin{equation}
\frac{Z_n}{Z^n}=\frac{\sum_rF_r^{nA}/{d_r^{2n-2}} }{(\sum_r F_r^A )^n}=\frac{1+\sum_{r\neq 0}c_r^{nA}/d_r^{2(n-1)}}{(1+\sum_{r\neq 
0}c_r^A)^n}.
\label{eq:zrat}
\end{equation}
The entanglement entropy then is
\begin{equation}
S_A=-\left.\frac\partial{\partial n}\frac{Z_n}{Z^n}\right|_{n=1}
=\log(1+\sum_{r\neq 0}c_r^A) -\frac{\sum_{r\neq0}c_r^A\log c_r^A/d_r^2}{1+\sum_{r\neq 0}c_r^A}.
\label{eq:sa}
\end{equation}
Note that that the series of character expansion coefficients is vanishing ($1>c_r>c_s| {\rm if}\, d_s>d_r$) in such a way that the sums in 
(\ref{eq:zrat}) are converging even for the smallest surface $A=1$. One then can choose the surface area large enough to guarantee that 
(\ref{eq:sa}) is finite.
We observe that this expression is $l$-independent\footnote{The independence on the size of the entangled region is a consequence of 
the absence of physical degrees of freedom in $2D$ gauge theory.} and valid for $l>0$. The entanglement entropy expression (\ref{eq:sa}) is 
universal in the sense that it does not depend on the initial lattice cutoff and is dependent only on the physical dimensions of the system. 
This is due to the fact that after a number of iterations the coefficients are attracted to the renormalization group (RG) trajectory 
independently of the starting point. It is interesting that  at $l=0$ $Z_n$ factors into n copies of $Z$ so that ratio $Z_n/Z^n=1$ and 
$S_A=0$. This is reminiscent  of the 2-dimensional theory end-point phase transition at temperature $T=0$.

If the surface area is very large one can truncate the series to obtain a manageable expression. This is in fact similar to the strong 
coupling limit treatment. Using strong coupling expansion in evaluation of $F_r$ we obtain an approximate expression for the 
entanglement entropy (\ref{eq:sa}).  In general one can compute $F_r$ term by term to any desired order. For our purposes, however, it is 
enough to keep the first two lowest order terms, which give the coefficients for  the trivial $r=0$ and fundamental $r=1$ representations 
\begin{equation}
F_r\approx\int dU \left(1+\frac\beta{2N} [ \chi_1(U) +h.c.] \right)\frac{1}{d_r}\chi^*_r(U).
\end{equation}
Thus $F_0=1$ and $c_1=F_1=\beta/(2N^2)$ for $N>2$ (note that characters of $SU(2)$ group are self-conjugate and therefore $c_1=
\beta/N^2$). The entropy becomes
\begin{eqnarray}
S_A&=&\log(1+(\frac\beta{2N^2})^A) -\frac{(\frac\beta{2N^2})^A\log \left((\frac\beta{2N^2})^A/N^2\right)}{1+(\frac\beta{2N^2})^A}\nonumber\\
&\approx&\left(\frac\beta{2N^2}\right)^A\left(1-\log \left((\frac\beta{2N^2})^A/N^2\right)\right).
\end{eqnarray}

Simplifications can be also achieved in the large $N$ limit. The expressions for the first two representations $F_r$ (integrals (\ref{eq:fr})) 
are readily available \cite{Gross:1980he}. In the Gross-Witten paper notation $F_0=z$ and $c_1=\omega$
\begin{equation}
F_1=\omega z=F_0\times
\left\{\begin{array}{ll}
1/\lambda,& \lambda\ge2\\
1-\lambda/4,&\lambda\le2
\end{array}\right. ,
\end{equation}
where $\lambda=g^2N$ is the `t Hooft coupling.

 Again for very large surface area it is reasonable to assume that the terms in this series are rapidly vanishing. Therefore the 
entanglement entropy becomes
\begin{equation}
S_A\approx \omega^A(1-\log \frac{\omega^A}{N^2}).
\end{equation}
Note that for the strong coupling $\omega=1/\lambda=\beta/(2N^2)$,  and the expression for $S_A$ is equal to the strong coupling 
expansion derived earlier (we can interchange strong coupling and large $N$). It is no surprise that the entanglement entropy is sensitive 
to the 2D Gross-Witten phase transition and is different for strong and weak coupling phases.

\subsection{$d\ge2$ gauge theory}
\begin{figure}[ht]
\centering
\includegraphics[width=0.55\columnwidth]{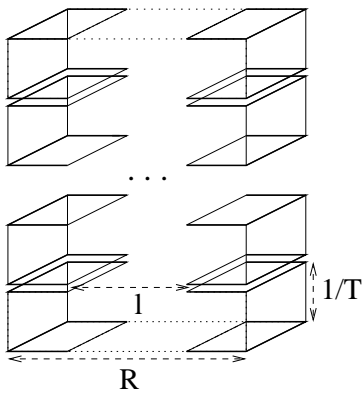}
\caption{$Z_n$ for $2+1$ dimensional theory.}
\label{fig:zn3d}
\end{figure}

Next we consider the $D=d+1$ dimensional theory, with $d\ge2$. This is a nontrivial theory which cannot be solved exactly. We employ the 
Migdal-Kadanoff (MK) \cite{Migdal:1975zg,Migdal:1975zf,Kadanoff:1976jb} decimation procedure to solve this theory approximately. For 
illustrative purposes we concentrate on $2+1$ theory, the generalization to higher dimensions is straightforward. 

In general for a finite temperature system one has to use 
anisotropic lattice. The time and space direction bond moving can be performed independently, c.f. $\lambda$- and $\rho$-
transformations for finite temperature gauge theory \cite{Bitar:1985hs,Imachi:1986pz,Imachi:1987ps}. To simplify the treatment we 
consider a vanishing temperature system in a symmetric box.

The standard MK decimation procedure ($\lambda$-transformation) moves  the internal plaquettes to the hyper-surfaces which constitute 
the elementary cells of the resulting coarse lattice
\begin{eqnarray}
\label{eq:MK}
e^{-S_p(U)}&=&\left[\sum_r F_r^Ad_r\chi_r(U)\right]^{\zeta^{1-b}},\\\nonumber
\quad F_r&=&\int dU e^{-\zeta^{b}S_p(U)}\frac{1}{d_r}\chi^*_r(U).
\end{eqnarray}
where the choice $b=0$ corresponds to Migdal, while $b=1$ to Kadanoff prescription. Here  
$\zeta=\lambda^{D-2}$ is the factor by which we strengthen the interaction on the resulting coarse lattice in order to compensate for 
missing internal plaquettes, $A=\lambda^2$ is the surface of the new elementary plaquette in units of fine lattice plaquettes (number of 
tiling fine plaquettes),  and
$\lambda$ is the scaling factor of the RG transformation and is equal to the number of plaquettes (internal and from the surface) moved to 
the surface from each of $D-2$ directions. 

It is known that the Kadanoff procedure results in the overcompensation of the strength of the coupling thus resulting in the upper bound 
for partition function, on the other hand leaving the coupling on the surface unchanged while dropping internal interactions $\zeta=1$ 
results in the lower bound on the partition function \cite{Tomboulis:2007iw}
\begin{equation}
Z(\zeta=1)\le Z\le Z(\zeta=\lambda^{D-2}).
\end{equation}
This relation relies on translation invariance and therefore does not hold for $Z_n$, however one may expect it to hold approximately. As 
a result a generalization of the MK procedure which preserves the partition function may be possible to construct. Here, however, we use 
the standard MK decimation.

After each step of decimation iteration the partition function decomposes into the product of  the coarse lattice partition function and the  
integrated out bulk part, which (after $m$ step iteration) is
\begin{equation}
\prod_{j=0}^mF_0(j)^{|\Lambda|/\lambda^{jD}}.
\end{equation}
Significant simplification can be achieved if we cary out decimations for $Z_n$ and $Z$ in exactly the same manner. 
As a result of equal volumes the bulk contributions in $Z_n$ and $Z^n$ are identical and cancel out in their ratio at each step. 

We start with a symmetric $D=d+1$ dimensional decimation ($\lambda$-transformation (\ref{eq:MK})) in $Z$ and $Z_n$, see Fig. 
\ref{fig:zn3d}. Note that there are periodicity conditions
in t-direction for each $\bar A$ part of n-replicas (bold links) and for the links of time boundary of the glued system ($Z_n$) belonging to 
$A$ (dotted links). The decimation should be altered
when the lattice spacing becomes equal to $l$ (the smallest scale in the problem). At this point
the $l$-like plaquettes (directed along $l$) inside the slab of thickness $l$ (extending through all $n$ replicas) have to be treated 
differently. These plaquettes can be decimated in remaining directions, very much like time-like plaquettes in the finite temperature 
gauge theory treatment. Such transformations are normally referred as  $\rho$-transformations, for them the decimation prescription 
(\ref{eq:MK}) is modified 
\begin{eqnarray}
\label{eq:MKl}
e^{-S_{p;l}(U)}&=&\left[\sum_r F_r^\lambda d_r\chi_r(U)\right]^{\zeta^{1-b}},\\\nonumber
\quad F_r&=&\int dU e^{-\zeta^{b}S_{p;l}(U)}\frac{1}{d_r}\chi^*_r(U).
\end{eqnarray}
We still can move plaquettes in $D-2$ direction but the tiling is done with $\lambda$ plaquettes.
All the other plaquettes are unaffected by this change and are decimated according to standard
($\lambda$-transformation) procedure. 

In appendix A we consider a gauge theory formulated in a box ${ R}^3$. This system will be the building block for construction of 
expressions for $Z_n$ and $Z$. 

Let us assume that the imaginary
surfaces that cut out the part for which we compute the entanglement entropy belong to $x-t$ planes and are
distance $l$ apart in $y$ direction (note that we consider 2+1).

We begin with $Z$. The surface with normal along $x$ consist of 3 pieces after decimation is stopped. There is 
exactly the same contribution from the surface with normal $-x$. 
At the center there is a boundary of the slab $c^s_x$ and two pieces which complement it, we refer to their joint as $\bar c^s_x$. 
The combined contribution is $c_x=c^s_x \bar c^s_x$ and should be substituted into the corresponding equation from the appendix. We 
note here that technically the complement to the slab (more precisely two complementary volumes) are not symmetric, therefore the 
recursion at some point has to be switched from $\lambda$ to $\rho$, however we consider the scale $R>>l$ and therefore we can 
always take $R$ large enough so that the corresponding coefficients are in the strong coupling limit and no transition from the RG flow to 
the infrared fixed points can occur in these bulks.

The 2 surfaces with normals $\pm y$ each contribute $c_y$. There is only one group of surfaces (similar to $c_x$) with normal 
$-t$ (our convention) with contribution $c_t=c^s_t\bar c^s_t$. Therefore from (\ref{eq:zcube})
\begin{equation}
Z=1+\sum_{i\neq0}(c^s_{x,i}\bar c^s_{x,i}c_{y,i})^2+\sum_{i,j\neq0}(c^s_{x,i}\bar c^s_{x,i}c_{y,i})^2d_j c^s_{t,j}\bar c^s_{t,j}D^i_{ij},
\end{equation}

After many successive steps of decimation iteration the only remaining degrees of freedom are defined on the surface of the system. In 
case of $Z_n$ we also have $n-1$ $l$-like plaquettes inside the bulk ($c^s_{t,j}$).
At this point we can move these $l$-like plaquettes in $Z_n$ to the bottom surface. This decimation step has no counterpart in the 
denominator and therefore the bulk term ($\tilde{F}_0$) of this last decimation procedure does not cancel. This decimation step is 
achieved only with moving the internal plaquettes along the time direction onto a single surface plaquette; there is no integration of the 
tiling plaquettes for this procedure, therefore the new coefficients for the resulting surface plaquette after Kadanoff type moving are
\begin{equation}
\tilde{F}^s_{t,j}=\int dU\left(1+\sum_{i\neq0}d_ic^s_{t,i}\chi_i(U)\right)^n\frac1{d_j}\chi_j(U^\dagger)
\end{equation}
and 
\begin{equation}
\tilde c^s_{t,j}=\frac{\tilde{F}^s_{t,j}}{\tilde{F}^s_{t,0}}.
\label{eq:ct_zn}
\end{equation}

Next we assume that the boundary between region $\bar A$, which has space-like links with coordinates $t=0$ and $t=1/T$ identified and 
region $A$ which has no such constraint is defined in such a way that the end links (directed along $x$) of the cut in $Z_n$ belong to 
region $A$. Since there is no periodicity requirement for these links we can integrate them out. As a result the internal $n-1$ time-like 
surface terms of $\bar A$ have support only at the trivial representation and therefore do not contribute to the partition function.
There is still, however, a contribution from the first replica time-like surface (bottom) of $Z_n$.

After simple considerations one can convince oneself that the surface integral in $Z_n$ is similar (in $2+1$ dimensional theory up to 
factor $1/d_i^{4(n-1)}$) to the surface integral of a $nN_t\times N_r^2$ cube\footnote{This seems to be a general property valid for any 
number of dimensions: the surface integral in $Z_n$ is equal to the surface integral in $Z$ with correspondingly increased volume up to 
an extra factor $1/d^{m(d)(n-1)}$, where $m(d)$ is some dimension dependent integer.}.
The side surface coefficients are modified to account for gluing $n$ replicas, while the bottom surface coefficient involves a term 
computed according to  (\ref{eq:ct_zn}) and is $\bar c^s_{t,i}\tilde c^s_{t,i}$. The partition function becomes
\begin{eqnarray}
&&Z_n\equiv \tilde F^s_{t,0} \cdot f_n=\tilde F^s_{t,0}
\times\\\nonumber
&&\left( 
1+\sum_{i\neq0}\frac1{d_i^{4(n-1)}}(c^s_{x,i}\bar c^s_{x,i}c_{y,i})^{2n}\left[1+\sum_{j\neq0}d_j \bar c^s_{t,j} \tilde c^s_{t,j}D^i_{ij}\right]
\right),
\end{eqnarray}

The ratio of the partition functions including the bulk term is
\begin{eqnarray}
&&\frac{Z_n}{Z^n}=\tilde F^s_{t,0}\times\\\nonumber
&&\frac{1+\sum_{i\neq0}(c^s_{x,i}\bar c^s_{x,i}c_{y,i})^{2n}/{d_i^{4(n-1)}}\left[1+\sum_{j\neq0} d_j \bar 
c^s_{t,j}\tilde c^s_{t,j}D^i_{ij}\right]}{\left(1+\sum_{i\neq0}(c^s_{x,i}\bar c^s_{x,i}c_{y,i})^2\left[1+\sum_{j\neq0}d_j c^s_{t,j} \bar 
c^s_{t,j}D^i_{ij}\right]\right)^n}
\end{eqnarray}
In order to obtain a higher dimensional expression for this ratio one needs to adjust accordingly the sides contribution and the 
contribution from the surface integration.

The entanglement entropy is
\begin{equation}
S_A=-\dot{\tilde F}^s_{t,0}+\log Z -\frac{\dot f_n}Z
\label{eq:sa_fin}
\end{equation}
where the dot stands for $\dot X=\left.\frac\partial{\partial n}X\right|_{n=1}$.
Note that
\begin{widetext}
\begin{equation}
\dot{f}_n=\sum_{i\neq0}(c^s_{x,i}\bar c^s_{x,i}c_{y,i})^{2}\log\frac{(c^s_{x,i}\bar c^s_{x,i}c_{y,i})^{2}}{d_i^4}\left(1+\sum_{j\neq0}d_j \tilde 
c_{t,j}D^i_{ij}\right)
+\sum_{i\neq0}(c^s_{x,i}\bar c^s_{x,i}c_{y,i})^{2}\sum_{j\neq0}d_j \bar 
c^s_{t,j} \dot{\tilde c}_{t,j}^sD^i_{ij}
\end{equation}
\end{widetext}
In order to manipulate these expressions we will need the following derivatives
\begin{eqnarray}
\dot{\tilde c}_{t,j}^s&=&\dot {\tilde{F}}_{t,j}^s-c^s_{t,j}\dot {\tilde{F}}_{t,0}^s\\
\dot {\tilde{F}}_{t,j}^s&=&\int dU [1+\sum_{i\neq0}d_ic^s_{t,i}\chi_i(U)] \log(1+\sum_{i\neq0}d_ic^s_{t,i}\chi_i(U))
\nonumber\\
&\times&\frac1{d_j}\chi_j(U^\dagger),
\end{eqnarray}

The expression for the entanglement entropy can be evaluated if the system flows towards the IR fixed point. This is the strong coupling 
limit for $c_{t,j}^s$ therefore we can expand logarithms and simplify the expression.
\begin{eqnarray}
\dot {\tilde{F}}_{t,j\neq0}^s&=&c^s_{t,j}+\sum_{i,i'\neq0}\frac{d_id_{i'}}{d_j} c^s_{t,i}c^s_{t,i'}D^j_{ii'}=c^s_{t,j}+O(c^2)
\nonumber\\\nonumber
\dot {\tilde{F}}_{t,0}^s&=&\sum_{i,j\neq0}c^s_{t,i}c^s_{t,j}d_id_j\int dU \chi_i(U)\chi_j(U)=O(c^2)
\end{eqnarray}

The leading term in the entropy is 
\begin{equation}
S_A\approx -(c^s_{x,1}\bar c^s_{x,1}c_{y,1})^{2}\log(c^s_{x,1}\bar c^s_{x,1}c_{y,1})^{2}
\end{equation}
Note that the dependance on $l$ is encoded in the value of $c^s_{x,1}$.

\subsection{Analyzing the RG flow}

Now recall that our choice of temperature makes the box symmetric and $c^s_{t,i}=c^s_{x,i}=c^s_i$. 
The resulting expression for the entanglement entropy (\ref{eq:sa_fin}) is a very complicated function of $c^s_i$. Note that this is a general 
feature valid for higher dimensional theories as well. The dependence on $l$ enters through the value of these coefficients. Essentially $l
$ regulates the moment when $\lambda$-transformation is switched to $\rho$-transformation, which in turn sets the initial value for the 
$c^s_{i}(m_0)$ iteration under $\rho$-transformations thus defining where the theory will flow before reaching the boundary.

\begin{figure}[ht]
\centering
\includegraphics[width=0.99\columnwidth]{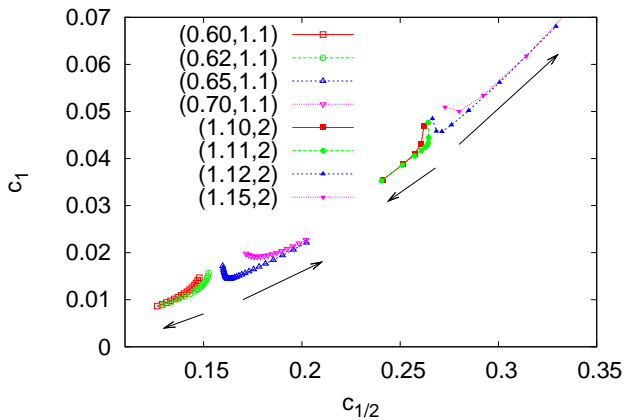}
\caption{Migdal decimation flow for $3+1$ dimensional $SU(2)$ gauge theory. Projection to  $c^s_{1/2}$ and $c^s_1$; $(\beta,\lambda)$ 
are indicated.}
\label{fig:flowm4d}
\end{figure}

Next we analyze the RG flow of $SU(2)$ gauge theory for $c^s_{i}(m)$ as a function of number of iterations $m$ under Migdal recursion 
(\ref{eq:MKl}) and depending on the starting point. In Fig. \ref{fig:flowm4d} we plot the projection of the flow (for a $3+1$ dimensional 
theory) from the infinite dimensional space of character coefficients onto the fundamental-adjoint $c_{1/2}-c_1$ plane. We consider $
\lambda=1.1$ and $2$ values and observe a significant dependence on the choice of the scaling factor. In what follows we will use the 
former value, since it is known to reproduces the $SO(3)$ critical coupling value \cite{Greensite:1981hw}. This value was also used to 
extract an approximately correct phase diagram for the mixed action fundamental-adjoint $SU(2)$ gauge theory \cite{Bitar:1982bp}.

One can clearly observe that depending on the starting value the flow will go to either of the two fixed points - the infrared trivial fixed point 
or non-trivial UV fixed point. This is a clear indication of a transition. The starting point for the system is set by the $\lambda$-
transformations and depends on the value of $l$. Generally at the starting point the action is a single plaquette action but with an infinite 
number of couplings for terms in all irreducible representations. In the numerical simulation that gives Fig. \ref{fig:flowm4d} we simplify this 
situation by considering a starting action in the wilsonian (only fundamental representation) form on $N_t=1$ lattice, noting that this 
should not affect the observed picture of existence of transition.

The lattice inverse coupling value $\beta^*_c\in(0.62,0.65)$ where the transition in the flow occurs should be compared to $N_t=1$ 
gauge theory finite temperature phase transition $\beta_c\approx0.86$ \cite{Velytsky:2007gj}. This allows one to relate the scale $l^*_c$ 
of the entanglement entropy transition to the finite temperature phase transition scale $l_c=1/T_c$. For this we use the standard 1-loop 
scaling relationship
\begin{equation}
a(\beta)\Lambda_L=\left(\frac\beta{2Nb_0}\right)^{b1/2b_0^2}\exp\left(-\frac\beta{4Nb_0}\right),
\end{equation}
where $b_0=11/24\pi^2$ and $b_1/2b_0^2=51/121$. Substituting the couplings we obtain
\begin{equation}
l^*_c/l_c\in(1.56,1.66).
\end{equation}

\section{Discussion of the results}
\begin{figure}[ht]
\centering
\includegraphics[width=0.95\columnwidth]{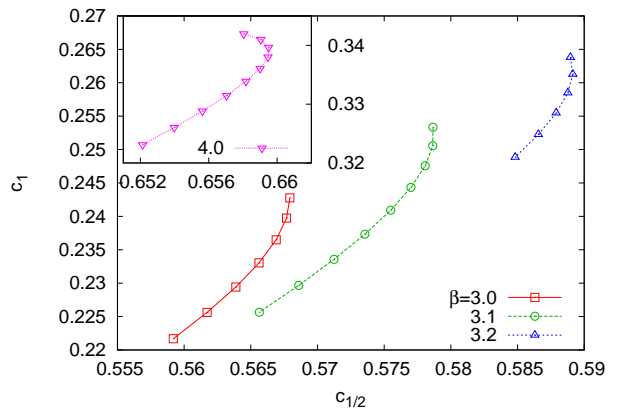}
\caption{Migdal decimation flow for $2+1$ dimensional $SU(2)$ gauge theory. Projection to  $c^s_{1/2}$ and $c^s_1$; $\lambda=1.1$, $\beta=3.0,3.1,3.2$ and $4.0$ (inlet). 
}
\label{fig:flowm3d}
\end{figure}
In this paper we studied the entanglement entropy in $d+1$ $SU(N)$ gauge theory. We use the multi-replica trick to relate the 
entanglement entropy to a simple ratio of partition funcitons. The $d=1$ theory is solved exactly. Free spatial b.c. lead to trivially zero 
entanglement entropy. Periodic spatial b.c. show non-zero universal value independent of the size $l$ of the entangled region. As the 
entangled region is removed the entropy becomes zero, showing in this manner 
behavior similar to the end point phase transition of $1+1$ dimensional theories.

Using MK decimation we approximately computed the ratio of partition functions and entanglement entropy in $d\ge2$ dimensional 
theories. A note of caution should be made regarding our choice to cary out the decimation for $Z_n$ and $Z$ in the same way. This 
allows us to significantly simplify the computational 
procedure. The non-analyticity in the RG flow observed for $Z_n$, however, 
 is also induced in $Z$ by this choice. This should not be a problem if one is interested only in the location of the transition.

In the case of $3+1$ $SU(2)$ gauge theory we demonstrated that there is a non-analytical change in the RG flow for coefficients of 
character expansions which define the entanglement entropy. We find that the length scale of this transition is $l^*_c/l_c\in(1.56,1.66)$.
Unfortunately the systematic error due to the use of the MK approximation is not easily tractable. 
It is interesting that in large $N_c$ case it was shown \cite{Klebanov:2007ws} that $l^*_c/l_c=2$. 

It is important to note that the MK procedure does not find a transition in the RG flow for $2+1$ dimensional theories. This transition is only 
observed for $d+1$ theories with $d>2$. Most likely this is an artifact of the MK procedure and $d=2$ theory exhibits a transition similar to 
higher dimensional theories. The MK decimation is known to miss the order of phase transition while correctly identifying its location. It is 
conceivable that in $d=2$ theory a proper transition can be seen by the MK procedure as a cross-over. This observation is supported by 
the fact that we indeed observe an interesting qualitative change in the flow around $\beta=3.2$, 
see Fig. \ref{fig:flowm3d}. For values of lattice inverse coupling 
below this value the flow is directed immediately toward the IR critical point (monotonously decreasing series of $c_i,\, \forall i$), while for 
larger values of $\beta$ the flow is directed from the IR fixed point for a few steps of iteration then switching to the flow toward the IR fixed 
point (initial increase of $c_{1/2}$ followed by monotonous decrease). We illustrate such a scenario 
for weak coupling regime $\beta=4.0$ in the inlet of Fig. \ref{fig:flowm3d}.
It is interesting that formally the MK $\rho$-transformation in $d+1$ theory with compact direction can be effectively viewed as $\lambda$-
transformation with an effective RG scaling parameter $\sqrt\lambda$ in $2D-2=2d$ dimensions, see 
\cite{Bitar:1985hs,Imachi:1986pz,Imachi:1987ps}. Thus $d>2$ theories are related to the zero temperature theories above the critical dimension 4 (have bulk phase transition), 
while $d=2$ is related to 4 dimensional zero temperature theory.

Similar results hold for $SU(3)$ and other $N_c$ groups. Therefore our claim is that the transition in the entanglement entropy is 
observed for any number of colors $N_c$ and the critical scale $l_c^*$ where transition takes place most likely is $N_c$ dependent and 
asymptotically reaches 2 as $N_c\rightarrow\infty$.

We note that the finite temperature phase transition studies of $SU(2)$ and $SU(3)$ gauge theory within MK formalism 
\cite{Bitar:1985hs,Imachi:1986pz,Imachi:1987ps} relied on the same analysis of the RG flow. It is important to emphasize that the periodic 
boundary conditions in time direction do not play any role in such studies. One has to impose the periodicity on the $N_t=1$ system after 
$\lambda$-transformations are switched to $\rho$-transformations \cite{Bitar:1985hs}. This will result in an 
effective lower dimensional spin system which exhibits a phase transition for $d\ge2$.

The study of the entanglement entropy effectively is transformed into an MK analysis of a gauge system defined with one compact 
direction and no periodicity imposed. Possibly Monte-Carlo simulations of such systems can define the location of the transition more 
accurately. However, this would be still a crude approximation since the MK treatment results in a well defined boundary, which is in 
reality  rather soft. Therefore direct numerical computation of the entanglement entropy should be preferred.

It is also interesting to relate our results to studies of the 
vortex free-energy order parameter \cite{tHooft:1979uj}, which 
provides a complete characterization of the possible phases of gauge theory.  
For $SU(2)$ it was found \cite{Kovacs:2000sy} that when the transverse 
size of the lattice is around $0.7fm$ there is a sharp cross-over in the vortex free energy. This cross-over has an obvious physical 
interpretation:  
the lattice size has to be large enough to accomodate sufficient spreading of the vortex flux ('fat' vortex) to enter the regime of exponential 
free-energy lowering by further spreading, i.e the 
confining or color magnetic mass-gap creation regime.   
Assuming that $\sqrt\sigma=420MeV$ ($\sigma$ is the string tension) we get in this theory $1/T_c=0.681fm$. Therefore the transition in the vortex free energy happens 
approximately at $1/T_c$ scale.

Using this observation we suggest that the transition in the entanglement entropy happens when the size of the entangled region is large 
enough to accommodate a fat vortex. The difference in the geometry should account on small difference of the scales when such 
transition occurs.

\section*{Acknowledgments}
The author would like to acknowledge insightful comments from D. Kutasov and T. Tomboulis.
This work was supported by the Joint Theory Institute funded together by 
Argonne National Laboratory and the University of Chicago.
This work is supported in part by the U.S. Department of Energy, Division of High Energy Physics and Office of Nuclear Physics, under Contract DE-AC02-06CH11357.

\section*{Appendix: $2+1$ dimensional gauge theory in a box}
\begin{figure}[ht]
\centering
\includegraphics[width=0.4\columnwidth]{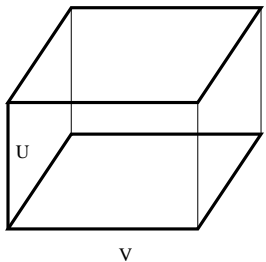}
\caption{$2+1$ dimensional symmetric box.}
\label{fig:cube}
\end{figure}
In this appendix we consider a $2+1$ dimensional $SU(N)$ gauge theory formulated 
in a symmetric box $R^3$ at temperature $T=1/R$, which corresponds to confined phase temperatures for sufficiently large $R$. The theory is formulated 
on a lattice with the UV cut-off $a$, with periodic boundary condition in $t$ direction and free boundary condition in spatial directions. The 
MK decimations (\ref{eq:MK}) with scale factor $\lambda$ are performed iteratively $N$ times ($\lambda^N=\hat R\equiv R/a$). At this 
point all degrees of freedom are "pushed" to the boundary and the resulting lattice spacing becomes equal to $R$, see Fig.
\ref{fig:cube}. Here we are not interested in the bulk contribution.
There are 8 independent gauge degrees of freedom which live on links. 
We use normals to the cube faces, which are directed outside to identify plaquettes. The partition function has contributions from 4 
plaquettes with normals along spatial directions and one plaquette (due to periodicity) from  $t$-direction 
\begin{equation}
f(\{c_{z}\}; \partial A)\equiv 1+\sum_{i\neq0}d_ic_{z;i}\chi_i(\partial A_z),
\end{equation}
where $z=\pm x,\pm y,t$ marks the plaquettes. The character coefficients $c_{z;i}$ can be obtained numerically and are the result of the 
RG flow in infinite dimensional coupling space. The symmetry of the box implies  $c_{x;i}=c_{y,i}=c_{t,i}$. For a general non-symmetric 
box one has to consider a series of $\rho$-transformations, resulting in an anisotropic lattice with all coefficients $c_{z,i}$ different.

Because of the free spatial boundary condition we can further integrate out three time-like links (thin lines in Fig. \ref{fig:cube}). By doing 
this we join the surfaces according to (\ref{eq:join}), with the resulting surface term
$f(\{c_{xy,i}\};U^\dagger V U V^\dagger)$ and $c_{xy,i}=c^2_{x,i}c^2_{y,i}$.
The partition function is
\begin{eqnarray}
Z&=&\int dU dV f(\{c_{xy,i}\};U^\dagger V U V^\dagger) f(\{c_{t,i}\};V)\nonumber\\
&=&1+\sum_{i\neq0}c_{xy,i}+\sum_{i,j\neq0}c_{xy,i}d_j c_{t,j}D^i_{ij},
\label{eq:zcube}
\end{eqnarray}
where
\begin{equation}
D^k_{ij}=\int dV \chi_k(V^\dagger)\chi_i(V)\chi_j(V)
\end{equation}
we recognize as the coefficients of the Clebsch-Gordan series ${\mathcal D}^{(i)}\times {\mathcal D}^{(j)}=\sum_k D^k_{ij} {\mathcal 
D}^{(k)}$ for the Kronecker product of irreducible representations. Using Gaunt's formula
\begin{eqnarray}
|G|^{-1}\int_G {\mathcal D}^{(j_1)}(R^{-1})_{n_1m_1} {\mathcal D}^{(j_2)}(R)_{n_2m_2} {\mathcal D}^{(j_3)}(R)_{n_3m_3}dR\nonumber\\
=
\left(
\begin{array}{c}
  j_1   \\
  n_1  \mu
\end{array}
\right)
\left(
\begin{array}{c}
  j_1   \\
  \nu m_1
\end{array}
\right)^*
\left(
\begin{array}{ccc}
  j_1& j_2  &j_3   \\
  \mu& n_2  & n_3  
\end{array}
\right)^*
\left(
\begin{array}{ccc}
  j_1& j_2  &j_3   \\
  \nu& m_2  & m_3  
\end{array}
\right),
\end{eqnarray}
where $|G|$ is the volume of the group space, we can express $D^k_{ij}$ through the Wigner coefficients (1-$j$ and 3-$j$ symbols) for 
general group \cite{Wigner:1959,Hamermesh:1962}
\begin{equation}
D^k_{ij}=
\left(
\begin{array}{c}
  k   \\
  n_1  \mu
\end{array}
\right)
\left(
\begin{array}{c}
  k   \\
  \nu n_1
\end{array}
\right)^*
\left(
\begin{array}{ccc}
  k& i  &j   \\
  \mu& n_2  & n_3  
\end{array}
\right)^*
\left(
\begin{array}{ccc}
  k& i  &j   \\
  \nu& n_2  & n_3  
\end{array}
\right).
\end{equation}

Coefficients $D^r_{rs}$ can be easily evaluated for $SU(2)$ group, using the Clebsch-Gordan equation
\begin{equation}
\chi_i\chi_j=\sum_{k=|i-j|}^{i+j}\chi_k.
\end{equation}
Thus the integral becomes
\begin{equation}
D^i_{ij}=\int dV \sum_{k=0}^{2i}\chi_k(V)\chi_j(V)=H_1(2i-j),
\end{equation}
where $H_1(x)$ is the Heaviside step function ($H_1(0)=1$). Therefore
\begin{equation}
Z_{SU(2)}=1+\sum_{i\neq0}c_{xy,i}+\sum_{i,j\neq0;j\le2i}c_{xy,i}d_j c_{t,j}
\end{equation}

\bibliographystyle{apsrev}
\bibliography{avref}

\end{document}